\documentstyle[12pt,aps]{revtex}
\def\beq{\begin{equation}}
\def\bar{\begin{eqnarray}}
\def\eeq{\end{equation}}
\def\ear{\end{eqnarray}}
\def\l{\lambda}
\begin{document}
\title{Electromagnetic Modes in Deformed Nuclei}
\author{J. Kvasil, R.G. Nazmitdinov $^{1}$, A. Mackova , M. Kopal}
\address{Department of Nuclear Physics, Charles University, 18000 Praha 8,
Czech Republic}
\author{N. Lo Iudice}
\address{Universit\`{a} di Napoli "Federico II" and INFN
I-80125 Napoli, Italy}
\author{V.O. Nesterenko}
\address{${^1}$ Bogoliubov Laboratory of Theoretical Physics,
JINR, 141980 Dubna, Russia}
\maketitle

\vskip 0.2cm

\begin{abstract}
A strength function method is adopted  to describe
a coupling between electric and magnetic modes of different
multipolarity. The collective vibrations are
analysed for a separable residual interaction in the framework of the
random-phase approximation. The coupling between $M2$ and $E1$
giant resonances is considered as an illustrative example.
\end{abstract}

\vskip 0.2cm

PACS number(s): 21.10.R, 21.60.E, 21.60.J

\vskip 0.5cm

\section{Introduction}

The probe of nuclear structure via electromagnetic interactions
provides important information on the nuclear wave function
and dynamics of nucleon-nucleon interactions.
The nuclear response upon different external fields is measured
by means of electromagnetic transitions between different
quantum states. The transition amplitude at the emission (absorption)
of a photon of a given multipolarity is proportional to
matrix elements of multipole moments. Since in nuclear photo-processes
the wave length $\lambda \approx 1/k$ of the photon  is larger then
the nucleus radius $R$ (the long-wave approximation $kR<<1$) the
multipole moments take the form \cite{BM69}
\bar
{\cal M}(E\lambda \mu) &=&
\sum_{i=1}^{A} e_{eff}^{(\lambda)} (i) r_{i}^{\lambda} Y_{\lambda \mu}(\hat
r_{i})
\nonumber \\
{\cal M}(M\lambda \mu) &=&
\frac{\mu_{N}}{2} \sqrt{\lambda(\lambda+1)} \sum_{i=1}^{A} \Big(
g_{S,eff}^{(\lambda)} (i)
\Big[ \sigma_{i} \otimes Y_{\lambda-1} (\hat r_i) \Big]_{\lambda \mu}
\Big( \vec \sigma_{(i)} .
\vec Y_{\lambda \mu}^{\lambda-1}(\hat
r_{(i)}) \Big)
\nonumber \\ && \
+ g_{l,eff}^{(\lambda)} (i) \frac{4}{\lambda+1}
\Big[ l_{i} \otimes  Y_{\lambda-1} (\hat r_i) \Big]_{\lambda \mu}\Big].
\Big( \vec l_{(i)}
. \vec Y_{\lambda \mu}^{\lambda-1}(\hat r_{(i)}) \Big) \Big)
r_{(i)}^{\lambda-1}
\label{81}
\ear
where $e_{eff}^{(\lambda)}(i)$ is the nucleon effective charge of multipolarity
$\lambda$, $g_{S,eff}^{(\lambda)} (i)$ and $g_{l,eff}^{(\lambda)} (i)$ are the
corresponding effective spin and orbital gyromagnetic ratios, respectively, and
$\mu_{N}=\frac{e\hbar}{2mc}$ is the nuclear magneton.
In this approximation the total decay probability from the quantum state
$|i>$ to the quantum state $|f>$ is determined as

\bar
T_{i\to f}(E_{\gamma}) =&& 5.49985*10^{22}\sum_{\lambda}
\Big[ \frac{E_{\gamma}}{(197.327)}\Big] ^{2\lambda +1}
\frac{2\lambda +1}{\lambda [(2\lambda +1)!!]^2}\nonumber\\
&&\Big[ B(E\lambda , i\to f) + 0.011064*B(M\lambda, i\to f)\Big]
\ear

where the reduced transition probability for the electromagnetic transition of
type $X$ ($X=E$ or $M$) and multipolarity $\lambda $ is defined
as
\beq
B(X\lambda;I_{i}K_{i}^{\pi_{i}}\nu_{i} \to I_{f}K_{f}^{\pi_{f}}\nu_{f}) =
\frac{1}{2I_{i}+1} \Big|\langle\! I_{f}K_{f}\nu_{f} \Vert {\cal
M}(X\lambda) \Vert I_{i}K_{i}\nu_{i} \rangle\Big|^{2}
\label{79}
\eeq

Here the $B(E\l,i\to f)$ is given in $e^2${\it fm}$^{2\l}$,
$B(M\l,i\to f)$ is given in $\mu_N^2${\it fm}$^{2\l-2}$,
the transition energy $E_{\gamma}$ is defined in {\it MeV}
and the total decay probability
$T_{i\to f}(E_{\gamma})$ is given in $1/s$.

At high excitation energy the density of nuclear states becomes so large
that the description in terms of transition probabilities between individual
states loses its practical sense.
Strength function techniques are known to represent a very valuable tool for
studying the electromagnetic properties of nuclei in energy regions
with high level density \cite{Sol,KN,Kv1}.
For example, defining the strength function as
\beq
\label{sf}
S(X\l, E_{\gamma})=\sum_{\nu}B(X\l,gr\to \nu)\delta(E_{\gamma}-\omega_{\nu}),
\eeq
we can describe the distribution of strength of the electromagnetic
excitation of the ground state over a range
of the available excitation energy $E_{\gamma}$.
Consequently, the strength function can be used for the analysis
of the photo-absorption cross
section
$\sigma (E_{\gamma})$
\bar
\label{crs}
\sigma (E_{\gamma})= &&1.40534*10^5 \sum_{\lambda}
\Big[ \frac{E_{\gamma}}{(197.327)}\Big] ^{2\lambda -1}
\frac{\lambda +1}{\lambda [(2\lambda +1)!!]^2}\nonumber\\
&&\Big[ S(E\lambda , i\to f) + 0.011064*S(M\lambda, i\to f)\Big]
\ear
Here, $S(E\lambda , i\to f)$, $S(M\lambda , i\to f)$
are given in $\frac{e^2{\it fm}^{2\l}}{MeV}$,
$\frac{\mu_N^2{\it fm}^{2\l -2}}{MeV}$, respectively.
The comparison of the strength functions extracted from
photo-absorption experiments with theoretical estimates
can testify the basic model assumptions
used for the calculations. Notice, that in the expression
Eq.(\ref{crs}) there is a restriction due to only the
parity selection rules.
Therefore, such excitations contain contributions from different electric
and magnetic operators which are able to create the excitation with a given
parity.

In axially deformed nuclei rotational bands are built on the
intrinsic states characterized by the angular momentum projection $K$
and the parity. In fact, such intrinsic states can be created by
different electric and magnetic operators with all possible
multipolarities in the nuclear Hamiltonian.
However, there is a restriction on the parity quantum number again.
For example, the state with $K^{\pi}=1^-$ can be created
by $E1, M2, E3, M4...$ operators. In standard approaches the coupling
between different electric and magnetic modes
allowed by the parity selection rules is neglected \cite{Sol,RS},
however, there is no justification for this assumption.
Based on the strength function method \cite{KN},
we propose the generalisation of this approach\cite{Kv1} which
allows consistently treat the contribution of different
electro-magnetic modes to the strength of excitations
(de -excitations), in particular, in region of giant resonances.

\section{The Model}

For the analysis of collective and single-particle degrees
freedom of nucleus we use the the phonon+rotor model described
in details in \cite{Kv1}. The model Hamiltonian is
\beq
\label{ham}
H=H_{\it sp}+H_{\it pair}+H_{\it res}+H_{\it rot}
\eeq
where $H_{\it sp}$ is a spherical one-body Hamiltonian
(a spherical Nilsson model), $H_{\it pair}$ represents
the pairing residual interaction, $H_{\it res}$
stands for the consistent with the one-body Hamiltonian
the long-range residual interaction and $H_{\it rot}$ is
the Hamiltonian of the rotor \cite{Sol}.
We use the ansatz of
separable multipole-multipole and spin-multipole-spin-multipole
forces including the isoscalar and isovector components
\bar
H_{\it res} &=& -\frac{1}{2} \sum \limits_{\lambda \mu}
\sum \limits_{\tau=0,1} \kappa_{\lambda} [\tau] \enspace
M^{\dagger}_{\lambda \mu}[\tau] M_{\lambda \mu}[\tau]
 - \frac{1}{2} \sum \limits_{l \lambda \mu}
\sum \limits_{\tau=0,1} \kappa_{l \lambda } [\tau]
\enspace
S^{\dagger}_{l \lambda \mu}[\tau] S_{l \lambda \mu}[\tau]
\ear
Here $\kappa_{\lambda} [\tau], \kappa_{l \lambda } [\tau]$
are the multipole strength constants and
\bar
\label{7}
M_{\lambda \mu}[^0_1] &=& M^{(p)}_{\lambda \mu} \pm M^{(n)}_{\lambda \mu},
\nonumber \\
S_{l\lambda \mu}[^0_1] &=& S^{(p)}_{l \lambda \mu} \pm
      S^{(n)}_{l \lambda \mu},
\ear
are isoscalar ($\tau = 0$) and isovector ($\tau=1$)
multipole fields, which are obtained from proton and
neutron multipole operators of the form:
\bar
\label{8}
M_{\lambda \mu}^{(\tau)} &=&
     \sum \limits_{^{q_{1} \sigma_{1} \in \tau}_
     {q_{2} \sigma_{2} \in \tau }}
     \langle q_{1} \sigma_{1} | R_{\lambda}(r) Y_{\lambda \mu}(\hat r)
     | q_{2} \sigma_{2} \rangle a^{\dagger}_{q_{1} \sigma_{1}}
     a_{q_{2} \sigma_{2}},
\nonumber \\
S_{l \lambda \mu}^{(\tau)} &=&
     \sum \limits_{^{q_{1} \sigma_{1} \in \tau}_
     {q_{2} \sigma_{2} \in \tau }}
     \langle q_{1} \sigma_{1} | R_{l \lambda}(r)
     [\sigma \otimes Y_{l}(\hat r)]_{\lambda \mu}
     | q_{2} \sigma_{2} \rangle a^{\dagger}_{q_{1} \sigma_{1}}
     a_{q_{2} \sigma_{2}},
\ear
For symmetry reasons it is convenient to construct multipole fields
of a good signature
\beq
\label{11}
R^{-1}_{1} F [r] R_{1} = r F [r]
\eeq
where $r=\pm1$ and $R_1\equiv e^{i\pi {\hat J}_x}$.
The details about the symmetry properties of the operators
can be found in \cite{Kv1}. After performing the signature
transformation, the Hamiltonian Eq.(\ref{ham}) can be expressed
through the quasiparticle creation (annihilation)
$\alpha^{\dagger}_{q\sigma} (\alpha_{q\sigma})$ operators
\beq
\label{ham1}
H=<HFB|H|HFB>+\sum \limits_{k} \varepsilon_{q} \sum \limits_{\sigma}
\alpha^{\dagger}_{q\sigma}\alpha_{q\sigma} + H_{\it pair}
+ H_{\it res} +  H_{\it rot}
\eeq
in the Hartree-Fock-Bogoliubov approximation.
Here $\varepsilon_q$ is the quasiparticle energy.
The pairing interaction
$H_{\it pair}$ and the long-range residual interaction $H_{\it res}$
expressed in terms of the quasiparticle operators consist of terms proportional
to combinations of type
$\alpha^{\dagger} \alpha^{\dagger}\alpha^{\dagger} \alpha^{\dagger}$,
$\alpha^{\dagger} \alpha^{\dagger}\alpha^{\dagger} \alpha$,
$\alpha^{\dagger} \alpha^{\dagger}\alpha \alpha$ and h.c..
At the description of even-even nuclei the terms proportional
to $\alpha^{\dagger} \alpha^{\dagger}\alpha^{\dagger} \alpha$ and h.c.
are not considered, since they create (annihilate) the odd number
of particles. The terms proportional to
$\alpha^{\dagger} \alpha^{\dagger}\alpha^{\dagger} \alpha^{\dagger}$ and
h.c are usually neglected in the description of the
vibrational states in the harmonic approximation
(see e.g. \cite{Sol,RS}).
The remaining terms proportional to
$\alpha^{\dagger} \alpha^{\dagger}\alpha \alpha$ are treated in the
random phase approximation (RPA). After solution of the RPA equation
of motion for collective modes \cite{KN} the Hamiltonian Eq.(\ref{ham1})
can be expressed in terms of the RPA creation and annihilation
phonons ${\cal Q}^{\dagger}_{\nu}$, $ {\cal Q}_{\nu}$
\beq
\label{rpa}
H_{RPA}=\sum \limits_{\stackrel{\nu}{\omega_{\nu} \not= 0}} \omega_\nu
({\cal Q}^{\dagger}_{\nu} {\cal Q}_{\nu} + \frac {1}{2}) + \frac{1}{2}
\sum
\limits_{\stackrel{\nu_{0}}{\omega_{\nu_{0}} = 0}} {\cal P}^{2}_{\nu_{0}},
\eeq
and the corresponding phonon energies $\omega_{\nu}$.
The last term in Eq.(\ref{rpa}) represents the contribution of the
spurious modes, for example, related to the rotation and translation
in the coordinate space or to the rotation in the isospin space \cite{KN,RS}.
The RPA phonon defined by the collective coordinate $\cal X_{\nu}$ and
momentum $\cal  P_{\nu}$ has the following two-quasiparticle structure
\beq
\label{Q}
{\cal Q}^{\dagger}_{\nu }= \frac{1}{\sqrt{2}}\Big[ \sqrt{\omega_{\nu}}
{\cal X}_{\nu}  - \frac{i}{\sqrt{\omega_{\nu }}} {\cal P}_{\nu }\Big]
=\sum_{k l}\left(\psi^{\nu }_{k \hat l} \alpha_{k {\hat l}}^{\dagger}
\alpha_{k {\hat l}}^{\dagger} -
\varphi_{k {\hat l}}^{\nu } \alpha_{{\hat l}k} \alpha_{{\hat l}k} \right)
\eeq
Notice that Eq.({\ref{Q}) is valid only for the positive signature phonons
which consist of the two-quasiparticle states
$\alpha_{k {\hat l}}^{\dagger} \alpha_{k {\hat l}}^{\dagger} $.
A similar expression can be written for the negative signature phonons
built from the two-quasiparticle states $\alpha_{k l}^{\dagger}
\alpha_{kl}^{\dagger}$ and $\alpha_{\hat k \hat l}^{\dagger}
\alpha_{\hat k \hat l}^{\dagger}$. Since in the phonon+rotor
model the rotation is treated adiabatically, there is a degeneracy
with respect to the quantum number of signature and all physical
properties can be described within the space of the positive signature
states only \cite{Kv1}. In this approximation the wave function
of the system can be written in the following form
\beq
|\nu IMK^{\pi}\rangle = \sqrt{\frac{2I+1}{16\pi^{2}(1+\delta_{K0})}}
\Bigl\lbrace
{\cal D}_{MK}^{I}(\theta) |\psi_{\nu \pi K}\rangle +
(-1)^{I} {\cal D}_{M-K}^{I}(\theta) R_{1} |\psi_{\nu \pi K}\rangle
\Bigr\rbrace
\label{Psitot}
\eeq
where $|\psi_{\nu \pi K}\rangle $ is the intrinsic wave function determined
by one-phonon or multi-phonon states.
The solution of the RPA equation of motion leads to the system of
algebraic equations which defines the characteristic
equation \cite{KN,Kv1}. The size of this equation is determined by the number
of the operators involved in the residual interaction.
It is clear that the phonon energies being the roots of this equation and the
structure of the two-quasiparticle amplitudes $\psi^\nu $ and $\varphi^\nu $
are  dependent on the type of the residual interaction. In the other words,
the presence or absence of the coupling between the electric and magnetic
modes should  influence on the structure of the RPA solutions.

\section{Discussion}

In order to avoid the cumbersome procedure of finding  all phonon
energies $\omega_{\nu}$ needed for the determination of the
strength function $S(X\lambda, E_{\gamma})$, Eq.(\ref{sf}), we developed the
averaging technique \cite{KN}. The advantage of this procedure consist of the
avoiding: i)the resolving the many-dimensional
characteristic equation for each single root $\omega_{\nu}$
and ii)calculations of corresponding eigenvectors of the Hamiltonian and
the $B(X\lambda, \omega_{\nu})$.
We consider the averaged strength function
\beq
\label{sf1}
{\cal S}_{\Delta}(\omega) = \sum_{\nu} B(\omega) \,
\rho_{\Delta} (\omega - \omega_{\nu})
\eeq
with the averaging function such
\beq
\rho_{\Delta} (x) = \frac{\Delta}{2\pi} \frac{1}{x^{2}+(\Delta /2)^{2}},
\qquad \int \limits_{-\infty}^{+\infty}\!\! dx\,\rho_{\Delta} (x) = 1 \  ,
\qquad \lim_{\Delta \to 0} \rho_{\Delta} (x) = \delta (x)
\eeq
In Ref.~\cite{Mal81} it was shown that the
results are fairly independent of the choice of the
averaging function.
Considering the function Eq.(\ref{sf1}) as a complex function
with a complex argument, we apply the Cauchy theorem and finally
obtain the following result
\bar
\label{sf2}
& &{\cal S}_{\Delta} (X; \nu_{gr} I_{i}=K_{i}=0 \to K^{\pi} I; ~\omega) =
- \frac{2}{\pi}
\mbox{\rm Im} \frac{\det(\hat B(z))}{\det(\hat D(z))}
\bigg|_{z=\omega+i\frac{\Delta}{2}} + \nonumber\\
& &\frac{\Delta}{2\pi} \sum_{ij} (p_{i\hat j}^{XIK})^{2} \bigg[ \frac{1}{\Big[
(\varepsilon_{i}+\varepsilon_{j}) - \omega\Big]^{2} + \frac{\Delta^{2}}{4}}
- \frac{1}{\Big[(\varepsilon_{i}+\varepsilon_{j}) + \omega\Big]^{2} +
\frac{\Delta^{2}}{4}} \bigg]
\ear

Here $D$ is the characteristic determinant of the RPA algebraic equations
and $p_{i\hat j}^{XIK}$ is a quasiparticle matrix element of the transition
operator. The matrix $B$ is similar to the matrix $D$, however,
it has the dimension by 1 greater than that of the matrix $D$
\bar
\begin{array}{cll}
B_{11}(\omega) = 0 \  , & B_{i1}(\omega) = - B_{1i}(\omega) \  ,
& (i=2, \dots, n) \\
\, & B_{ij}(\omega) = D_{i-1,j-1}(\omega) \  , & (i,j=2, \dots, n)
\end{array} \label{107}
\ear
The first column and the first row of the matrix $B$ contain sums
of the product of quasiparticle  matrix elements
of the transition operator and of the operators of the residual
interaction (see details in \cite{Kv1}).
The first term in Eq.(\ref{sf2}) is related to the residual
interaction and the second term is determined by the
mean field.

As the example of the coupling of different multipoles and spin-multipoles
in the Hamiltonian we consider the $M2$ strength
function ${\cal S}_{\Delta}(M2,E)=\sum \limits_K
{\cal S}_{\Delta}(M2,E; gr \to 2^-K)$ for $^{154}Sm$. In Fig.1
the solid line corresponds to the coupling of $M_{1K}$ (electric dipole)
and $S_{12K}$ (magnetic quadrupole) operators. The dashed line
represents the contribution of the magnetic quadrupole term alone.
The contribution of the plain quasiparticle term (the second term
in Eq.(\ref{sf2})) is shown with the dotted line. Whereas the residual
interaction changes the distribution of the strength of the magnetic quadrupole
transitions, the influence of the electric dipole
mode to the total strength function is relatively small.
It seems that in the region of giant resonances the coupling
between electric and magnetic operators for the M2 mode
is essentially not important.
The analysis of this coupling for low-lying states and for the
other operators is in progress.

Figure Capture

{\bf Fig.1} The total $M2$ strength function vs the excitation
energy $E$.

\begin{thebibliography}{99}
\bibitem{BM69} A. Bohr and B.R. Mottelson, {\it Nuclear Structure},
V.1 (Benjamin, New York, 1969)
\bibitem{Sol} V.G. Soloviev, {\it Theory of Atomic Nuclei:
Quasiparticles and Phonons} (Institute of Physics, Bristol, 1992)
\bibitem{KN} J. Kvasil and R.G. Nazmitdinov,
Sov. J. Part. Nucl. {\bf 17}, 265 (1986)
\bibitem{Kv1} J. Kvasil, N. Lo Iuduce, V.O. Nesterenko and
 M. Kopal,  Phys.Rev. {\bf C58}, 209 (1998)
\bibitem{RS} P. Ring and P. Schuck, {\it The Nuclear-Many Body Problem}
(Springer, New York, 1980)
\bibitem{Mal81} L.A. Malov, Preprint JINR, P4-1-228, 1981.
\end{thebibliography}
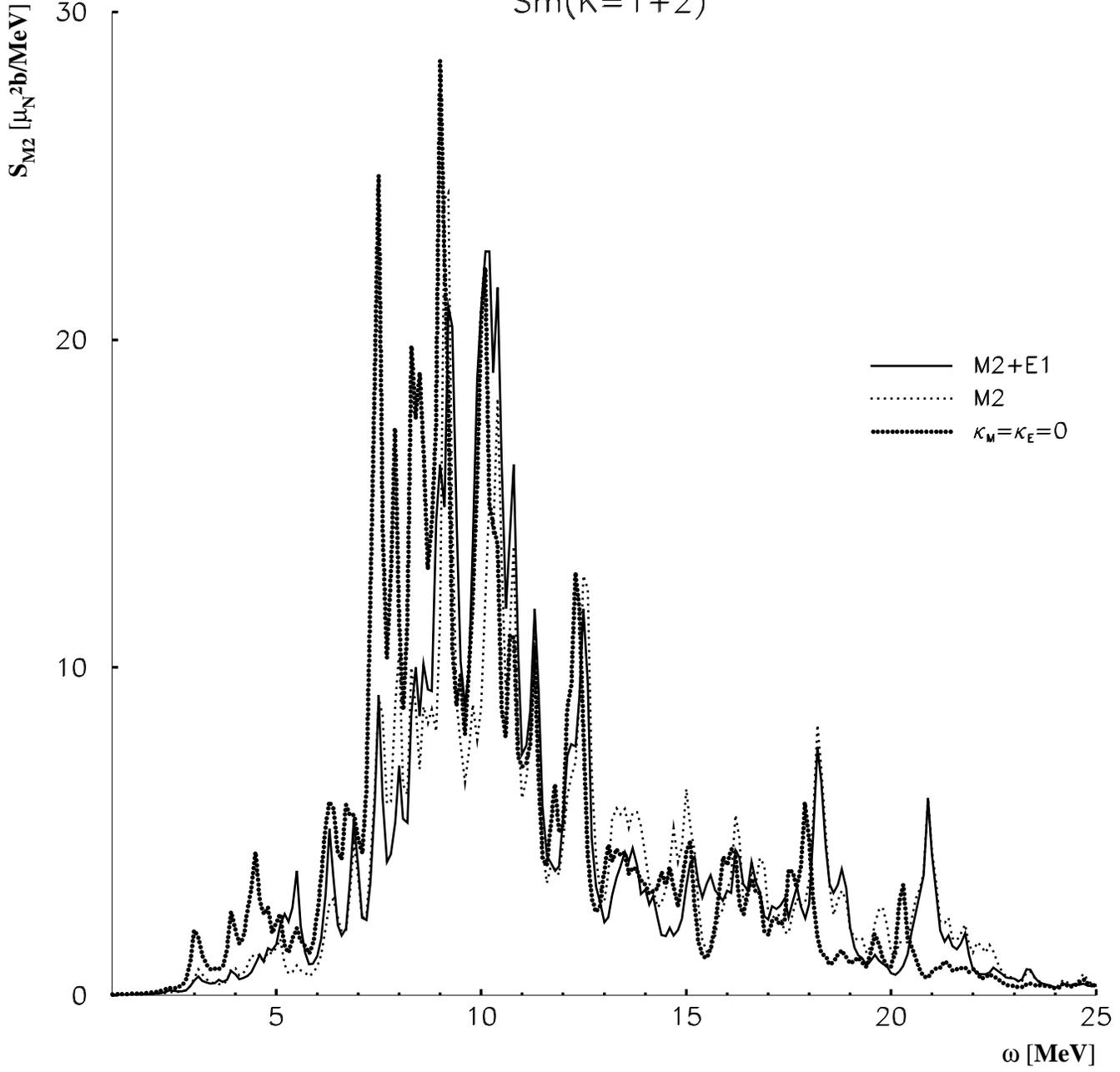
\end{document}